\documentclass[aps,groupedaddress,floatfix,nofootinbib,multirow,twocolumn]{revtex4}

\input epsf

\begin{document}
\title{
   Isospectral But Physically Distinct:\\
   Modular Symmetries and their Implications for Carbon Nanotori}
\author{Keith R. Dienes$^{1,2,3}$\footnote{E-mail address:  {\tt dienes@physics.arizona.edu}},
        Brooks Thomas$^{2,3}$\footnote{E-mail address:  {\tt brooks@physics.arizona.edu}}}
\affiliation{
     $^1$ Physics Division, National Science Foundation, Arlington, VA  22230  USA\\
     $^2$  Department of Physics, University of Maryland, College Park, MD  20742  USA\\
     $^3$ Department of Physics, University of Arizona, Tucson, AZ  85721  USA}
\begin{abstract}
   Recently there has been considerable interest in the properties of carbon
   nanotori.  Such nanotori can be parametrized according to their radii,
   their chiralities, and the twists that occur upon joining opposite ends
   of the nanotubes from which they are derived.  In this paper, however, we demonstrate
   that many physically distinct nanotori with wildly different parameters nevertheless share 
   identical band structures, energy spectra, and electrical conductivities.
   This occurs as a result of certain geometric symmetries
   known as modular symmetries 
   which are direct consequences of the properties of
   the compactified graphene sheet. 
   Using these symmetries, we show 
   that there is
   a dramatic reduction in the number of spectrally 
   distinct carbon nanotori compared with the number of physically distinct 
   carbon nanotori. The existence of these modular symmetries also allows 
   us to demonstrate that many statements 
   in the literature concerning the electronic properties 
   of nanotori are incomplete because they fail to respect the spectral equivalences 
   that follow from these symmetries.
   We also find that 
   as a result of these modular symmetries,
    the fraction of spectrally distinct nanotori which 
    are metallic is approximately three times greater than would naively be
    expected on the basis of standard results in the literature.
    Finally, we demonstrate that these modular symmetries also extend to cases in which
    our carbon nanotori enclose non-zero magnetic fluxes.
\end{abstract}
\maketitle

\newcommand{\newc}{\newcommand}
\newc{\gsim}{\lower.7ex\hbox{$\;\stackrel{\textstyle>}{\sim}\;$}}
\newc{\lsim}{\lower.7ex\hbox{$\;\stackrel{\textstyle<}{\sim}\;$}}

\def\beq{\begin{equation}}
\def\eeq{\end{equation}}
\def\beqn{\begin{eqnarray}}
\def\eeqn{\end{eqnarray}}
\def\calM{{\cal M}}
\def\calV{{\cal V}}
\def\calF{{\cal F}}
\def\half{{\textstyle{1\over 2}}}
\def\quarter{{\textstyle{1\over 4}}}
\def\threehalf{{\textstyle{3\over 2}}}
\def\ie{{\it i.e.}\/}
\def\eg{{\it e.g.}\/}
\def\etc{{\it etc}.\/}
\def\Nhex{N_{\mathrm{hex}}}


\def\inbar{\,\vrule height1.5ex width.4pt depth0pt}
\def\IR{\relax{\rm I\kern-.18em R}}
 \font\cmss=cmss10 \font\cmsss=cmss10 at 7pt
\def\IQ{\relax{\rm I\kern-.18em Q}}
\def\IZ{\relax\ifmmode\mathchoice
 {\hbox{\cmss Z\kern-.4em Z}}{\hbox{\cmss Z\kern-.4em Z}}
 {\lower.9pt\hbox{\cmsss Z\kern-.4em Z}}
 {\lower1.2pt\hbox{\cmsss Z\kern-.4em Z}}\else{\cmss Z\kern-.4em Z}\fi}

\input epsf



\section{Introduction}

Soon after the experimental discovery~\cite{Iijima} of carbon nanotubes,
it was suggested~\cite{Dunlap} that there might also exist 
carbon {\it nanotori}\/ --- 
\ie, carbon nanotubes in which the ends of the tube are identified and sewn together.
Within a few years, experimental evidence for such structures emerged~\cite{NanotorusDiscoveryLiu,NanotorusDiscoveryMartel},
and since then they have attracted a great deal of 
experimental~\cite{otherexptobservations,otherexptobservations2}
and theoretical~\cite{Haddon,PersistentMagMoments,ColossalMagMoments,MagneticPropertiesArmchair,
ExperimentalNanotoriAreTwisted,MarganskaSzopa,resonance,KaneMele,DoThetwist,Kirby,
SasakiKawazoe,CurvatureDistortion,strain,ThermalDisorder} attention.
There are a variety of reasons for this intense interest.
For example, certain species of carbon nanotori exhibit unusual magnetic
properties~\cite{Haddon,PersistentMagMoments,ColossalMagMoments,MagneticPropertiesArmchair},
including persistent magnetic moments at nearly zero flux~\cite{PersistentMagMoments} 
and colossal paramagnetic moments~\cite{ColossalMagMoments}.  These objects can 
also display a diverse 
variety of electric properties:   some nanotori are inherently metallic,
while others are semiconducting and still others are insulators.  

As we shall discuss, the most general nanotorus can be 
parametrized by four integers $(m,n,p,q)$.  Together and in various
combinations, these describe the radius of the underlying
nanotube, the chirality of the underlying nanotube, the length of the underlying
nanotube, and the relative twist~\cite{ExperimentalNanotoriAreTwisted,MarganskaSzopa}
that might occur upon sewing opposite ends of
the tube together to form the torus.
The important point, however, is that nanotori with different $(m,n,p,q)$ are
fundamentally physically distinct:   they have different sizes, different shapes, 
and different twisted ``honeycomb'' patterns of
carbon atoms laid out on their surfaces.
Of course, there are certain symmetries of the underlying graphene sheet which lead
to trivial equivalences amongst these nanotori.  For example, $60^\circ$ rotations
of the underlying graphene sheet will produce identical nanotori.  Such nanotori
are therefore not physically distinct.

In this paper, however, we shall demonstrate that there are additional symmetries
that relate {\it physically distinct}\/ nanotori to each other, forcing such nanotori to 
exhibit identical energy spectra and electrical properties. 
These so-called ``modular'' symmetries therefore
transcend the traditional hexagonal lattice symmetries of the graphene sheet,
and arise solely in the process of 
compactifying the graphene sheet in order to form the nanotorus.
In some sense, the appearance of these modular symmetries is entirely expected, for
they result directly from the geometry of the compactification.
However, to the best of our knowledge, the 
significance and consequences of these symmetries have not been fully appreciated
thus far in the nanotorus literature.

As we shall demonstrate, these symmetries have a number of
profound effects. 
First, we shall see that
use of these symmetries allows us to partition the 
set of carbon nanotori into distinct equivalence
classes as far as their spectral properties are concerned,
and leads to a dramatic reduction in the numbers of {\it spectrally distinct}\/ 
carbon nanotori as compared with the numbers of {\it physically distinct}\/ nanotori.
Moreover, as we shall show, many of the standard rules of thumb advanced in the 
literature in order to describe the conductivity
properties of these nanotori actually fail to respect these symmetries.
Such rules of thumb 
are therefore incomplete as descriptions of the physics of these
nanotori, and must be replaced by statements which respect the full symmetry structure of
the compactified graphene sheet.
Finally, we demonstrate that these symmetries 
even extend to situations in which  
the carbon nanotori enclose different types of magnetic flux.
They thus should have applicability for many of the fascinating magnetic
properties of carbon nanotori, including the possibility of persistent currents.

\section{Preliminaries:  ~The graphene sheet, the carbon nanotube, and the carbon
nanotorus}

In order to explain the origins of these spectral symmetries,
we begin with a brief review which will also serve to
highlight our notation and conventions.

\begin{figure}[t!]
\centerline{
   \epsfysize 3.0 truein \epsfbox{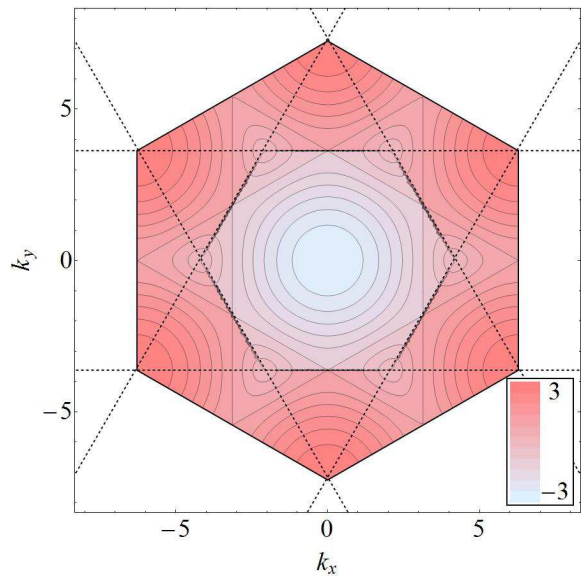} 
 }
\caption{Contour plot of the rescaled dispersion relation
$E(\vec k)/\gamma_0$ of the 
infinite graphene sheet for $\vec k$ within the first Brillouin zone 
and expressed in units of $R_{cc}^{-1}/\sqrt{3}$.
The dotted lines indicate the lowest-order Bragg ``planes'', and
the six points with $E=0$ at which these planes have pairwise intersections
constitute the corresponding Fermi ``surface''.} 
\label{fig:EnergyContours}
\end{figure}

In general, the graphene sheet   
is nothing but a set of carbon atoms arranged on an
extended, two-dimensional hexagonal lattice 
generated by two basis vectors $\vec a_1$ and $\vec a_2$.
We shall choose a Euclidean coordinate system such that 
$\vec a_1=(1,0)$ and $\vec a_2= (\half, -\half \sqrt{3})$
in units of $\sqrt{3}R_{cc}$, where $R_{cc}$ is
the fundamental carbon-carbon bond length.
As always, the band structure associated with a given lattice can be
described in terms of a dispersion relation $E(\vec k)$ which relates an electron
wavevector $\vec k$ to its energy $E$.
For the graphene sheet, and in 
the tight-binding (or H\"uckel) approximation in which the only 
significant overlap integrals are those
between the $2p_z$ orbitals associated with nearest-neighbor carbon
atoms, this dispersion 
relation is given by~\cite{resonance,KaneMele,DoThetwist}
\begin{eqnarray}
  E(\vec{k}) &=& 
     \pm \gamma_0\, \biggl[ 3 
         + 2\cos[\vec{k}\cdot (\vec{a}_1-\vec{a}_2)] \nonumber\\ 
    &&~~~~~~~ + 2\cos(\vec{k}\cdot \vec{a}_2) + 2\cos(\vec{k}\cdot \vec{a}_1) \biggr]^{1/2}~.
\label{eq:BandsTightBinding}
\end{eqnarray} 
Here $\gamma_0 \approx 0.266$~eV is the energy-transfer resonance integral 
between two neighboring $2p_z$ orbitals.  
A contour plot of this dispersion relation, clearly indicating the band
structure within the first Brillouin zone,
is shown in Fig.~\ref{fig:EnergyContours}.  

For the uncompactified graphene sheet, 
all electron wavevectors $\vec k$ are allowed. 
However, this situation changes when 
we ``roll up'' one dimension of the
graphene sheet to produce a single-walled carbon nanotube,
or equivalently when we 
identify any two carbon atoms on the graphene sheet whose
positions differ by an arbitrary lattice vector
$\vec V_1 = m \vec a_1 + n \vec a_2=(m+\half n, -\half \sqrt{3} n)$ with $(m,n)\in \IZ$.
Imposing the Bloch condition on the electron wavefunctions $\psi(\vec r)$
in addition to the new periodicity condition 
$\psi(\vec r) = \psi(\vec r+\vec V_1)$  
then restricts $\vec{k}=(k_x,k_y)$ to the set of wavevectors satisfying the condition  
\begin{equation}
         k_x L_1 \cos\beta+ k_y L_1 \sin\beta ~=~2\pi \ell_1 ~,~~~~~~~ \ell_1\in\IZ~,
\label{eq:tubek}
\end{equation} 
where $L_1^2\equiv |\vec V_1|^2 = m^2 + mn+n^2$
and $\cos \beta \equiv (m+\half n)/L_1$.
These allowed values of $\vec{k}$ therefore form parallel lines
in the $(k_x,k_y)$ plane, and the locations at which these lines intersect the 
Bragg planes in Fig.~\ref{fig:EnergyContours} determine whether the 
corresponding $(m,n)$ nanotube is metallic, semiconducting, or insulating.
Nanotubes for which $n=0$ (\ie, $\beta=0$) or $m=n$ (\ie, $\beta=\pi/6$)
are dubbed ``zigzag'' or ``armchair'' respectively;
nanotubes with other values of $(m,n)$ are generally referred to as ``chiral''.
In general, the vector $\vec T$ perpendicular to $\vec V_1$ is the tube axis.

\begin{figure}[b!]
\centerline{
   \epsfysize 2.15 truein \epsfbox{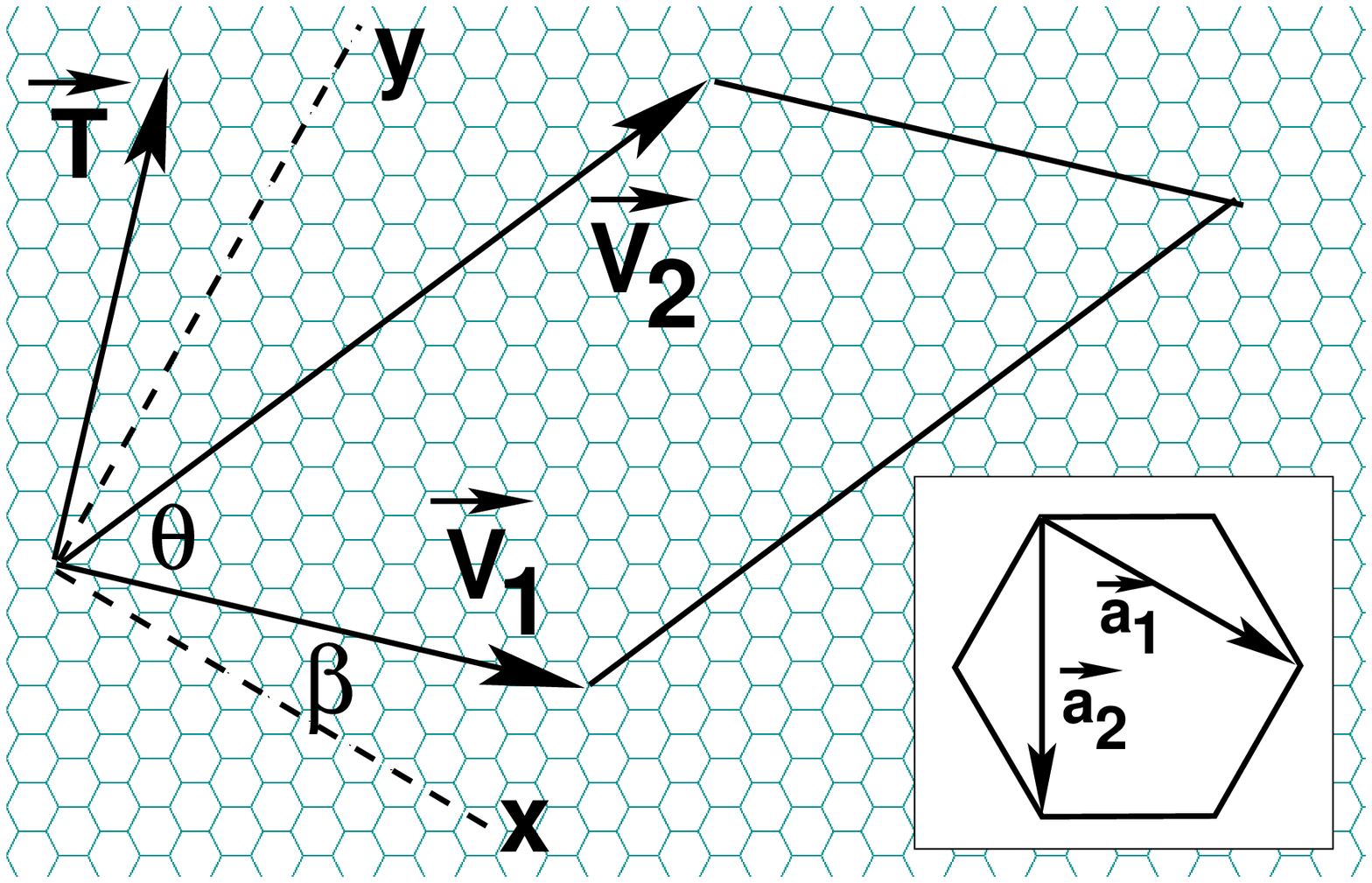} 
 }
\caption{Coordinate system $(x,y)$, lattice vectors $\vec a_{1,2}$, 
compactification vectors $\vec V_{1,2}$,
and tube axis vector $\vec T$ on the graphene sheet.}
\label{fig:TorusParam}
\end{figure}  

We now consider imposing {\it two}\/ 
non-parallel identifications on the graphene sheet,
as illustrated in Fig.~\ref{fig:TorusParam}.
In general, we consider two arbitrary lattice identification vectors 
$\vec{V}_1=m \vec a_1+ n \vec a_2$ and 
$\vec{V}_2 = p \vec a_1 + q \vec a_2$;
without loss of generality we shall assume that $-\pi/3<\beta \leq \pi/3$. 
We shall also assume without loss of generality that 
the quantity $N_{\rm hex}\equiv np-mq$ is positive, or
equivalently that the relative angle $\theta$ from $\vec V_1$ 
to $\vec V_2$ lies in the range $0<\theta<\pi/2$, 
as illustrated in Fig.~\ref{fig:TorusParam}. 
Such identifications together
result in a so-called $(m,n,p,q)$ nanotorus which may be viewed as
a $(m,n)$ nanotube in which opposite ends are joined to each other, 
potentially with a relative twist angle $\theta$.  
This description is especially appropriate if 
$L_2^2\equiv |\vec V_2|^2 = p^2 + pq+q^2 \gg L_1^2$.
Note that $N_{\rm hex}$ describes the number of hexagonal   
cells which tile the surface of the resulting donut, and  
is equal to precisely half the number of carbon atoms in the nanotorus. 
Imposing the double
periodicity condition 
$\psi(\vec{r}) = \psi(\vec{r}+\vec{V}_1) = \psi(\vec{r}+\vec{V}_2)$ 
on the electron wavefunction $\psi(\vec r)$
in conjunction with the Bloch condition
then leads to the constraint in Eq.~(\ref{eq:tubek})
as well as the additional constraint 
\begin{equation}
         k_x L_2 \cos\delta + k_y L_2 \sin\delta ~=~2\pi \ell_2 ~,~~~~~~~ \ell_2\in\IZ~,
\label{eq:tubek2}
\end{equation}
where $\delta\equiv \theta+\beta$.
Solving Eqs.~(\ref{eq:tubek}) and (\ref{eq:tubek2}) simultaneously
then yields
\begin{eqnarray}
  k_x &=& \frac{2\pi}{L_1L_2\sin\theta}
    \left[\ell_1 L_2\sin\delta -\ell_2 L_1\sin\beta\right] \nonumber\\
  k_y &=& \frac{2\pi}{L_1L_2\sin\theta}
    \left[-\ell_1 L_2\cos\delta +\ell_2 L_1\cos\beta\right] 
\label{eq:torik}
\end{eqnarray}
or equivalently
\begin{eqnarray}
  k_x &=&  {2\pi \over N_{\rm hex}} \left[  -\ell_1 q + \ell_2 n \right]\nonumber\\
  k_y &=&  {2\pi \over \sqrt{3} N_{\rm hex}} \left[ -\ell_1 (2p + q) + \ell_2 (2m+n) \right]~.
\label{eq:torik2}
\end{eqnarray}
Thus, we see that the allowed electron wavevectors now form a two-dimensional grid of points in the 
$(k_x,k_y)$ plane.  
Note that there are always 
exactly $N_{\rm hex}$ allowed wavevectors lying inside the fundamental
region bounded by the Bragg planes in Fig.~\ref{fig:EnergyContours}, where points lying on a
single Bragg plane itself are counted as half and where points lying at the intersections 
of two Bragg planes are counted as a third. 

Whether such nanotori are metallic, semiconducting, 
or insulating depends on whether these points hit or come particularly 
close to the intersections of the Bragg planes in Fig.~\ref{fig:EnergyContours}. 
It is found that nanotori in which both $m-n$ and $p-q$ are multiples of three
are metallic, with solutions to Eqs.~(\ref{eq:torik}) and (\ref{eq:torik2}) that lie
precisely on the Fermi surface.  
For practical applications, it proves useful to focus on only 
those nanotori with $L_2\gg L_1\gg R_{cc}$, as this 
condition allows one to bend the nanotube into a nanotorus without excessive strain or deformation
of the underlying graphene sheet (the effects of which would otherwise distort 
the dispersion relation in Fig.~\ref{fig:EnergyContours}, and hence the band structure
of the torus).
Within this limit, it is then conventional to regard nanotori with $m-n=0$~(mod~3) 
but $p-q\not=0$~(mod~3) as semiconducting, since the allowed wavevectors come ``close'' to the 
Fermi surface in this limit.  All other nanotori are then considered insulating.

\section{Modular symmetries and spectral equivalences of the carbon nanotorus}

All of the above results are completely standard, and are well known in the 
carbon-nanotorus literature.
In particular, tori with different values of $(m,n,p,q)$ are physically distinct: 
they have entirely different arrangements of hexagons tiling their surfaces,
with different values of the physical radii $L_{1,2}$, chiral angle $\beta$, and 
twist angle $\theta$.
We shall take this to be our definition of ``physically distinct''.
There are, of course, certain trivial identifications which relate 
different nanotori to each other:  for example, the $(m,n,p,q)$ torus 
and the $(m+n, -m, p+q,-p)$ torus are actually identical,
since they correspond to $(\vec V_1,\vec V_2)$ pairs which are related 
to each other by a uniform $60^\circ$ rotation.  
Such trivial identifications reflect
the underlying hexagonal lattice symmetries of 
the graphene sheet from which these nanotori are constructed, 
and result in identical carbon nanotori with identical 
patterns of carbon atoms on their surfaces.  Such nanotori are therefore not physically distinct. 

By contrast, an important question is whether there exist physically {\it distinct}\/ nanotori 
(\ie, tori which are {\it not}\/ related by symmetries of the hexagonal lattice)
which nevertheless yield identical grids of allowed wavevectors $(k_x,k_y)$.  If so,
we will have found cases of physically distinct tori with different values 
of $(L_1,L_2,\theta,\beta)$ which are nevertheless 
spectrally identical.  In other words,  such tori will have identical 
energy spectra and electrical conducting properties.

At first sight, it might appear that no such spectral equivalences exist for carbon nanotori.
After all, the results in Eq.~(\ref{eq:torik}) for carbon 
nanotori may initially appear to be nothing more 
than a two-dimensional generalization of the results in Eq.~(\ref{eq:tubek}) for carbon nanotubes,
and no such spectral equivalences exist for carbon nanotubes. 
However, it is important to realize that
this is not true for carbon nanotori.
In particular,  it turns out that the set of allowed values 
of $(k_x,k_y)$ in Eq.~(\ref{eq:torik2}) is actually invariant under two 
additional symmetry transformations
which we shall denote $S$ and $T$:
\begin{equation}
   S:~ \left\{
   \begin{array}{ccl}
     m&\to&-p\\
     n&\to&-q\\
     p&\to&m\\
     q&\to&n~,
   \end{array}
   \right.
  ~~~~~~~
   T:~ \left\{
   \begin{array}{ccl}
     m&\to&m\\
     n&\to&n\\
     p&\to&p+m\\
     q&\to&q+n~.
   \end{array}
   \right.~
\label{eq:STonIndices}
\end{equation}
Under these transformations, it is straightforward to demonstrate
that $N_{\rm hex}\equiv np-mq$ is invariant, while the physical
parameters $L_1$, $L_2$, $\theta$, and $\beta$ transform in the following manner:  
\begin{eqnarray}
   S:&&~ \left\{
   \begin{array}{ccl}
     L_1&\to& L_2\\
     L_2&\to& L_1\\
     \theta&\to&\pi-\theta\\
     \beta&\to&\beta+\theta-\pi
   \end{array}
   \right. \nonumber\\
   ~    \nonumber\\
   T:&&~ \left\{
   \begin{array}{ccl}
     L_1&\to& L_1\\
     L_2&\to& \sqrt{L_1^1+L_2^2+2L_1L_2\cos\theta}\\
     \cot\theta&\to& \cot\theta+ (L_1/L_2)\csc\theta\\ 
     \beta&\to&\beta
   \end{array}
   \right.
\label{eq:STonVars}
\end{eqnarray}
It is important to stress that the individual equations in Eqs.~(\ref{eq:torik}) 
and (\ref{eq:torik2})
are not invariant under $S$ or $T$;  rather, what is invariant is the {\it set}\/ 
of values of $(k_x,k_y)$ to which these equations lead. 
Moreover, since these sets of solutions for $(k_x,k_y)$ are invariant under $S$ and $T$
individually,
they are also invariant under any sequence of $S$ and $T$ transformations.
For example, under the $ST^{-1}ST(TS)^2$ transformation we find
$(m,n,p,q)\to (3m-2p,3n-2q,2m-p,2n-q)$, and this too is a symmetry
of the solutions to Eq.~(\ref{eq:torik2}).

It then follows from these observations
that any two tori whose defining vectors $\vec V_1$ 
and $\vec V_2$ differ through $S$ and $T$ transformations
share the same electronic spectra but are 
intrinsically different from each other --- \ie, that they are isospectral
but physically distinct.  That they are isospectral 
follows from the fact that the same set of $(k_x,k_y)$ solutions are selected
in each case.
By contrast, that they are physically distinct
follows from the fact that
the fundamental identifications between carbon atoms
on the graphene sheet
are altered by $S$ and $T$ transformations in a manner that transcends
trivial hexagonal lattice symmetries.
This is perhaps easiest to see in the case of the $T$ transformation, which 
corresponds to the action $(\vec V_1,\vec V_2)\to (\vec V_1,\vec V_1+\vec V_2)$.
This changes not only $L_2$ but $\theta$,
and thus produces a new torus which has a greater ``twist'' when the 
ends of the nanotube are joined.

\begin{figure}[t!]
\centerline{
   \epsfxsize 3.5 truein \epsfbox{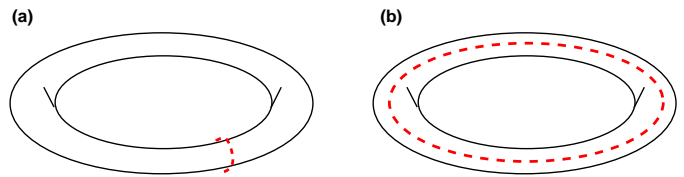} 
 }
\caption{Sketch of (a) the $T$ transformation, and (b) the $T'$ transformation,
as actions on the carbon nanotorus.  In each case, one should imagine affixing
the carbon atoms to the surface of the torus, then cutting the torus along
the dashed (red) line (thereby breaking those carbon/carbon bonds which cross
this line), twisting along the cut by a full $2\pi$ rotation, 
and then re-attaching the newly adjacent bonds to reconstruct a new nanotorus.  
The original and final nanotori will be related by the $T$ or $T'$ 
modular transformations respectively, while much more complex spectral equivalences 
can be generated through combinations
of these two fundamental operations.}
\label{tori}
\end{figure}

It may be less obvious that the $S$ transformation also connects physically
distinct tori.  Indeed, this transformation corresponds
to the action $(\vec V_1,\vec V_2)\to (-\vec V_2, \vec V_1)$,
and at first glance it might appear that this is merely a trivial relabeling
of the two independent periodicities, along with a reflection (sign flip).
However, 
we must remember that 
the $S$ transformation also changes the associated $\theta$ and $\beta$
angles in non-trivial ways.
Alternatively, 
we can also appreciate the non-trivial nature of $S$ by considering
the combination
\beq
           T' ~\equiv~ S T^{-1} S^{-1}~=~ T S T
\label{Tprimedef}
\eeq
which corresponds to the action $(\vec V_1,\vec V_2)\to (\vec V_1+\vec V_2,\vec V_2)$.
[The second equality in Eq.~(\ref{Tprimedef}) follows as a 
result of the identity $(ST)^3= -{\bf 1}$,
or equivalently $(T^{-1} T')^3= -{\bf 1}$.]
Note that $T'$ is in some sense ``dual'' to $T$:   each
performs a full twist around a different cycle of the torus. 
In other words, as illustrated in Fig.~\ref{tori}, while $T$ corresponds 
to cutting through one side of the
torus and re-attaching the two edges along with a twist, $T'$ corresponds to 
cutting the torus along the other cycle and then twisting along the cut
in an analogous manner before attaching the newly adjacent bonds.
Thus, $T'$ results in a physically distinct torus, just as $T$ does,
and furthermore these resulting tori are not related to each other through 
hexagonal lattice symmetries. 
It then follows from Eq.~(\ref{Tprimedef}) that $S$ also cannot correspond to a lattice
symmetry --- \ie, $S$ must connect physically distinct tori as well.
Indeed, $S$ and $T'$ are interchangeable in the sense that the two generators
of the modular group can be considered to be either $\lbrace S,T\rbrace$ or 
$\lbrace T,T'\rbrace$.

Note that if we define the complex quantity $\tau\equiv (L_2/L_1) e^{i\theta}$,
then the $S$ and $T$ transformations correspond to $\tau\to -1/\tau$ and $\tau\to \tau+1$ 
respectively.  Together, these transformations generate the so-called 
``modular group'',   
which is one of the primordial symmetries associated with 
toroidal compactifications.  
In general, any 
transformation which takes the form
$\tau\to (a\tau+b)/(c\tau+d)$ where $a,b,c,d\in\IZ$ and $ad-bc=1$
is a modular transformation.
What we have shown, then, is that 
any two carbon nanotori whose defining parameters
are related by a modular transformation
are spectrally identical even though they are physically distinct. 
In other words, any transformation which can be generated through repeated actions of the $S$
and $T$ generators leads to a spectral equivalence between 
physically distinct tori.


It is important to emphasize that the appearance of these modular symmetries
is not a total surprise.  Modular symmetries often
arise in the presence of toroidal compactifications, and are a direct consequence
of the underlying geometry of the carbon nanotorus.
Indeed, the zone-folding of a macroscopic lattice
necessarily introduces periodic equivalences of the sort discussed here, and in the
case of two dimensions with non-parallel identifications, such periodicities 
generically lead to modular symmetries. 
In other words, the underlying toroidal graph is not altered
by modular symmetries, and thus the adjacencies of the graph
(and the isospectrality which results) are preserved intact.
However, when applied to the case of carbon nanotori, these
symmetries are {\it physical}, in the sense that the nanotori they relate
are isospectral but physically distinct.
As we shall see, this 
is particularly significant, 
allowing these symmetries to have
profound effects on the physics
of these nanotorus systems and in particular on the range of behaviors they may exhibit.

In a sense, these modular symmetries can be viewed as a generalization
of a similar set of symmetries that emerge 
for M\"obius graphs.
To see this, let us imagine taking a long strip of hexagons and gluing
opposite ends of this strip together with an arbitrary number of half-twists.
At first glance, one might suspect that the resulting spectrum
should depend on the actual number of half-twists.
However, it turns out that the resulting spectrum depends only on whether
this number is even or odd~\cite{Mobius}.
This M\"obius case is, of course, a relatively trivial example of this phenomenon,
since it deals with only a single possible kind of twist 
corresponding to the single periodicity that defines the M\"obius strip.
Indeed, for the case of carbon nanotori,
our point is that there exists a whole
 {\it group}\/ of transformations --- the modular group --- which 
is built upon the actions of {\it two}\/ non-commuting underlying generators ($S$ and $T$) which can be 
combined and performed in a variety of inequivalent ways.
This leads to ``sewing'' and ``gluing'' configurations 
on the torus which are much more complex than on the M\"obius strip, and which require greater mathematical
machinery to study --- and this in turn leads to distinctions between the notions
of ``physically distinct'' and ``spectrally
distinct'' which are even more pronounced than in the M\"obius case.
However, even the relatively simple M\"obius case illustrates one of our basic themes: 
there can be lots of physically distinct twisted hexagon strips,
yet very few of them are spectrally distinct.
It is the preservation of the underlying graph in each case which makes this
possible.

As a result of these modular symmetries, we can henceforth establish an unambiguous
convention for uniquely describing 
the spectral properties of a given carbon nanotorus:
we calculate its complex parameter $\tau$ as defined above,
and then 
use a sequence of $S$ and $T$ modular transformations as needed in order
to bring $\tau$ into a special region of the complex
$\tau$-plane known as the ``fundamental domain'' ${\cal F}$
defined by 
\beq
  {\cal F}\/~\equiv~\lbrace \tau:  ~-\half < \tau_1 \leq \half~, ~\tau_2 >0~,~ |\tau|\geq 1\rbrace
\eeq
where $\tau_1\equiv {\rm Re}\,\tau$ and $\tau_2\equiv {\rm Im}\, \tau$.
Note that the condition $N_{\rm hex}\equiv np-mq >0$
already ensures that $\tau_2 >0$.
We can then use the underlying symmetries of the hexagonal graphene lattice
in order to bring the resulting angle $\beta$ into the range $0\leq \beta < \pi/3$.
Note that this convention is tantamount to describing the spectrum 
of a given carbon nanotorus in terms of that spectrally equivalent 
nanotorus which is as close to being rectangular as possible.  
Following this procedure therefore provides an unambiguous test of whether 
any two physically distinct carbon nanotori in fact have 
the same spectral properties.


\section{Physical implications}

The existence of these modular symmetries and the spectral equivalences
they induce has profound implications for the physics of carbon nanotori.
Indeed, as we shall see, the use of these modular symmetries will 
sharpen our ability to classify nanotori on the basis of those physical properties 
(such as their metallicities, {\it etc.}\/) which ultimately stem from 
their band structures.  Furthermore, use of these modular symmetries 
can even provide {\it statistical}\/ insights into the properties of randomly produced
carbon nanotori --- insights which may ultimately have practical consequences for
the manufacture of such objects in environments in which their resulting properties
often cannot be controlled or engineered in advance.

\begin{figure}[t!]
\centerline{
   \epsfysize 3.0 truein \epsfbox{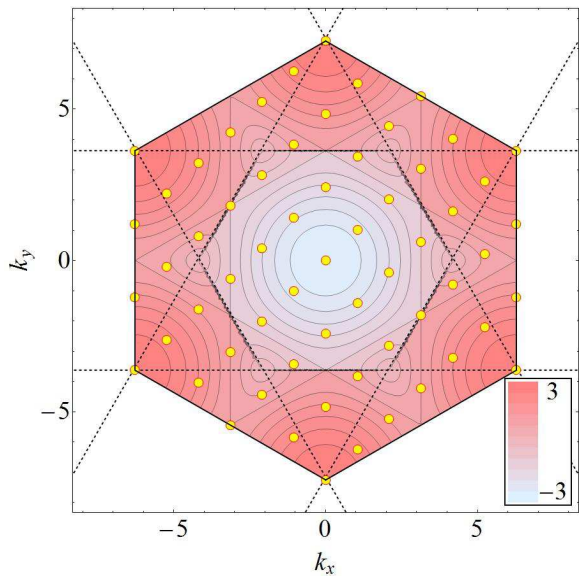} 
 }
\caption{A plot of the energy spectrum common to the 
     $(3,2, 24,10)$,  $(7,3,-22,-12)$, and $(8,6,-25,-21)$
    carbon nanotori, with allowed wavevectors indicated by (yellow) dots
    superimposed over the energy contours in Fig.~\protect\ref{fig:EnergyContours}. 
    Note that these nanotori are not metallic, 
    as none of the allowed wavevectors coincide with
    any of the six points which constitute the Fermi ``surface''.  
\label{fig:EnergySpecThreeTori}}
\end{figure}

As discussed above,
the first and most direct implication of these modular symmetries is that
there exist physically distinct nanotori ---  often with 
markedly different radii and chiral angles --- which nevertheless
possess identical energy spectra and conductivity properties.
As an example of this phenomenon, let us consider the 
$(3,2, 24,10)$,  $(7,3,-22,-12)$, and $(8,6,-25,-21)$
carbon nanotori.  
Clearly, we see that 
these nanotori are physically distinct and are characterized by 
different sets of cycle lengths $L_{1,2}$, different 
chiral angles $\beta$, and different twist angles $\theta$.  
As a result, these three tori have entirely different patterns of carbon atoms
tiling their surfaces. 
Yet, the latter two nanotori are in fact related to each another by $S$ and $T$
transformations, and consequently they each yield the same
spectrum of allowed $\vec{k}$ vectors shown in Fig.~\ref{fig:EnergySpecThreeTori}.   
Similarly the first nanotorus is related to the second two through not only $S$ and $T$
transformations, but also trivial 60$^\circ$ rotations of the underlying graphene sheet.  
Thus, these tori all share identical spectra, and
according to the convention specified above, we can describe this
spectrum uniquely as having $N_{\rm hex}=18$, $\tau =(2+ 9\sqrt{3}i)/13$,
and $\tan\beta= \sqrt{3}/7$.

Second, as a corollary to this observation, we also learn
that the usual notions of ``zigzag'' and ``armchair'' --- notions
which are critical for describing the chirality and metallicity of carbon 
nanotubes --- 
no longer hold as special indicators of the spectral properties   
of corresponding carbon nanotori.
In other words, nanotori built from zigzag or armchair nanotubes
might have spectral properties which are identical to those of nanotori built
from non-zigzag or non-armchair nanotubes;  moreover, these spectral properties
might or might not correspond to the special zigzag or armchair angles $\beta=0,\pi/6$.
As an example, 
the $(6,0,-17,-6)$ nanotorus is built from a zigzag nanotube
while the $(9,9,-40,-44)$ nanotorus is built from an armchair nanotube
and the $(3,15,13,53)$ and $(4,8,21,33)$ nanotori are built from chiral nanotubes
with different chiralities.
Yet all four nanotori have identical spectral properties
which are the same as those of a chiral nanotorus with 
$N_{\rm hex}=36$, $\tau= 12(1 + \textstyle{{3\over 2}}\sqrt{3}i)/31$, and $\tan\beta= 5\sqrt{3}/7$
(or $\beta \approx 51.05^\circ$).
Indeed, using the mathematical results of Ref.~\cite{Kirby}, we can show
that any carbon nanotorus is spectrally equivalent to one built from a zigzag carbon nanotube.

Conversely, a carbon nanotorus can have a spectrum which exhibits a zigzag or armchair property
that is lacking in the original nanotube from which it is constructed.
As an example, 
the $(9,0,-25,-4)$ nanotorus is built from a zigzag nanotube
and the $(4,7,-12,-30)$ nanotorus is built from a chiral nanotube.
Yet both nanotori have the spectral properties
of an armchair nanotorus, with $N_{\rm hex}=36$, $\tau= 3\sqrt{3}i/2$, and $\beta= \pi/6$.
Likewise, 
the $(9,9,-27,-31)$ nanotorus is built from an armchair nanotube and 
the $(3,6,-11,-34)$ nanotorus is built from a chiral nanotube.
Yet both nanotori have the spectral properties
of a zigzag nanotorus, with $N_{\rm hex}=36$, $\tau= (3+9\sqrt{3}i)/8$, and $\beta= 0$.

Third, it turns out that not every possible spectral signature
$(N_{\rm hex}, \tau, \beta)$ can be realized, even in principle.
Instead, these three quantities experience internal constraints and correlations 
which ultimately reflect the
fixed hexagonal lattice structure of 
the underlying graphene sheet and which require
that any allowed spectral signature $(N_{\rm hex}, \tau, \beta)$ 
originate from four real integers $(m,n,p,q)$.
These correlations amongst
$(N_{\rm hex}, \tau, \beta)$ 
can take a variety of different forms.
For example, for odd values of $N_{\rm hex}$,  
it turns out that there exist no self-consistent spectral signatures with 
purely imaginary values of $\tau$.
Similarly, for $N_{\rm hex}=24$, there are solutions with
${\rm Re}\,\tau\in \lbrace \pm 1/2, \pm 1/3, \pm 1/4, \pm 1/7\rbrace$, but no other
inverse integers;  likewise, all of these 
except for ${\rm Re}\,\tau=1/7$ have $\beta=0$.
In fact, these internal constraints amongst the four real parameters
embodied in $(N_{\rm hex},\tau,\beta)$ are sufficiently strong that they
effectively eliminate one real degree of freedom from within this parametrization. 
In other words, although four real degrees of freedom are required in order
to parametrize a given physically distinct nanotorus, only three real degrees
of freedom are required in order to uniquely describe its spectral properties~\cite{Kirby}. 

It is easy to understand why such constraints arise.
Ordinarily, as a question of topology, tori can exist with all shapes
and volumes, for  there are literally an infinite number of ways in which we can 
construct a torus by rolling up an unmarked sheet of paper.
However, in the case of carbon nanotori,
our original ``sheet of paper'' 
is not unmarked:  it is actually a graphene sheet of carbon atoms which has its own hexagonal
lattice structure.  The existence of such a lattice 
structure has a number of critical consequences:
it forces our toroidal defining vectors $\vec V_1$ to $\vec V_2$
to be {\it lattice vectors}\/; 
it restricts the resulting possible combinations of $N_{\rm hex}$ and $\tau$ to 
those values consistent with the periodicity of the lattice;  
and it breaks the rotational symmetry of 
our original uncompactified two-dimensional sheet and 
necessitates the introduction of a new measurable parameter,
the angle $\beta$ defined in Fig.~\ref{fig:TorusParam}, 
which describes the orientation of the torus relative to the underlying lattice.
Conversely, the presence of the hexagonal lattice gives a clear 
meaning to the twist angle $\theta$.  This quantity would have had no meaning
when rolling up an unmarked sheet of paper.

Fourth,
it is clear that 
the existence of spectral equivalences
between different nanotori
implies that the number of {\it spectrally}\/ distinct 
nanotori in any set
will necessarily be smaller than the number of 
 {\it physically}\/ distinct nanotori in that set.
However, it turns out that 
the magnitude of this truncation can easily become quite staggering.
As an example, let us consider 
the set of physically distinct $(m,n,p,q)$ nanotori 
with fixed $N_{\rm hex}=np-mq= 18$ which can be formed
from integers in the range $|m|,|n|,|p|,|q|\leq \Lambda$ for some
cutoff $\Lambda$.  For concreteness, we shall take $\Lambda=100$.
In order to count only physically distinct nanotori, we shall require
that $(m,n)$ be chosen such that $-\pi/3 < \beta \leq \pi/3$.
We shall also require, as a rough measure of their physical consistency in three-dimensional
space, that each such nanotorus 
have an outer circumference $L_2$ which is at least 
triple its inner cross-sectional circumference $L_1$; 
note that other similar mathematical conditions may alternatively be imposed, 
but the qualitative results to follow are essentially unchanged.
Given these constraints, we then find through direct enumeration 
that there are exactly $12\,205$ physically distinct carbon nanotori which have $N_{\rm hex}=18$.
Yet, as a result of these spectral equivalences,
it turns out that
these $12\,205$ physically distinct carbon nanotori 
give rise to only $14$ distinct energy spectra!
These distinct energy spectra are listed in Table~\ref{Nhex18table},
along with a representative carbon nanotorus in each class.

\begin{table}[t!]
\begin{center}
\begin{tabular}{||ccc|c|c||}
                 \hline
                 \hline
   $\tau_1$ & $\tau_2$ &  $\tan\beta$ & ~metal?~ & ~sample $(m,n,p,q)$~ \\ 
                 \hline
                 \hline
   0 & $\sqrt{3}$ &  0 & yes &  $( 12,  15,  42,  51)$ \\ 
                 \hline
   0 & $3\sqrt{3}$ & $\sqrt{3}/3$ & yes &  $( 10,  13, -34, -46)$ \\ 
                 \hline
   0 & $9\sqrt{3}$ & 0 & no &  $( 10,   8, -39, -33)$ \\ 
   0 & $9\sqrt{3}$ & $\sqrt{3}/27$  & no &  $( 19,  17, -66, -60)$ \\ 
\hline
  $\phantom{+}1/3$ & $\sqrt{3}$ & 0 & no & $( 10,  -6, -33,  18)$ \\
  $-1/3$ & $\sqrt{3}$ & 0 & no & $( 10,  -4,  33, -15)$\\ 
\hline
  $\phantom{+} 1/4$ & $3\sqrt{3}/4$ & $\sqrt{3}/3$ & yes & $( 10,  16, -32, -53)$\\ 
  $-1/4$ & $3\sqrt{3}/4$ & $\sqrt{3}/3$ & yes & $( 10,   1, -32,  -5)$\\ 
                 \hline
  $\phantom{+} 1/4$ & $9\sqrt{3}/4$ & 0 & no & $(10,  -1,  32,  -5)$\\
  $-1/4$ & $9\sqrt{3}/4$ & 0 & no & $(10,  18, -34, -63)$\\
\hline
  $\phantom{+}3/7$ & $9\sqrt{3}/7$ & $\sqrt{3}/2$ & no & $(10,  14,  32,  43)$\\ 
  $-3/7$ & $9\sqrt{3}/7$ & $\sqrt{3}/5$ & no & $(10,  12,  34,  39)$\\ 
\hline
  $\phantom{+}2/13$ & $9\sqrt{3}/13$ & $\sqrt{3}/7$ & no & $( 10,  11, -32, -37)$\\ 
  ~$-2/13$~ & ~$9\sqrt{3}/13$~ & ~$3\sqrt{3}/5$~ & no & $(10,  14, -33, -48)$\\
                 \hline
                 \hline
\end{tabular}
\end{center}
\caption{For $N_{\rm hex}=18$ and $\Lambda=100$, there are $12\,205$ physically distinct
carbon nanotori $(m,n,p,q)$ with $L_2 >3 L_1$.  However, these exhibit only 14 spectrally
distinct energy spectra and band structures, and only four of these correspond to metals.
These 14 spectrally distinct values of $\tau\equiv \tau_1+i\tau_2 = |\tau|e^{i\theta}$ 
and $\beta$ are listed above, along with a sample $(m,n,p,q)$ nanotorus in each class.}   
\label{Nhex18table}
\end{table}

This is clearly a major truncation.
Even with the relatively small value $N_{\rm hex}=18$ and
relatively small cutoff $\Lambda=100$,
each spectral signature listed in Table~\ref{Nhex18table} is shared
by literally hundreds or thousands of physically distinct carbon nanotori.

Fifth,
it is also worth noting that
this truncation does not treat metallic and non-metallic
nanotori equally.
In general, a given $(m,n,p,q)$ carbon nanotorus will be metallic if at least one 
of its allowed wavevectors $\vec k$ lies on the Fermi surface, or in this case
on one of the six points at which two Bragg planes intersect.
As is well known, this occurs only when the differences 
$m-n$ and $p-q$ are each a multiple of three.
This implies that in any large set of physically distinct carbon nanotori,
approximately one-ninth of the nanotori should be metallic.
However, we see from Table~\ref{Nhex18table} that four out of the fourteen possible distinct
spectral signatures are metallic.  This is almost triple what would have been expected,
implying that the fraction of spectrally distinct nanotori which are metallic is  
nearly triple the fraction of physically distinct nanotori which are metallic. 
In other words, the truncation from the four-parameter space of physically distinct
nanotori to the three-parameter space of spectrally distinct nanotori
is remarkably sensitive to the metallicity of the nanotori in question.

\begin{figure*}[thb!]
\centerline{
   \epsfxsize 3.4 truein \epsfbox{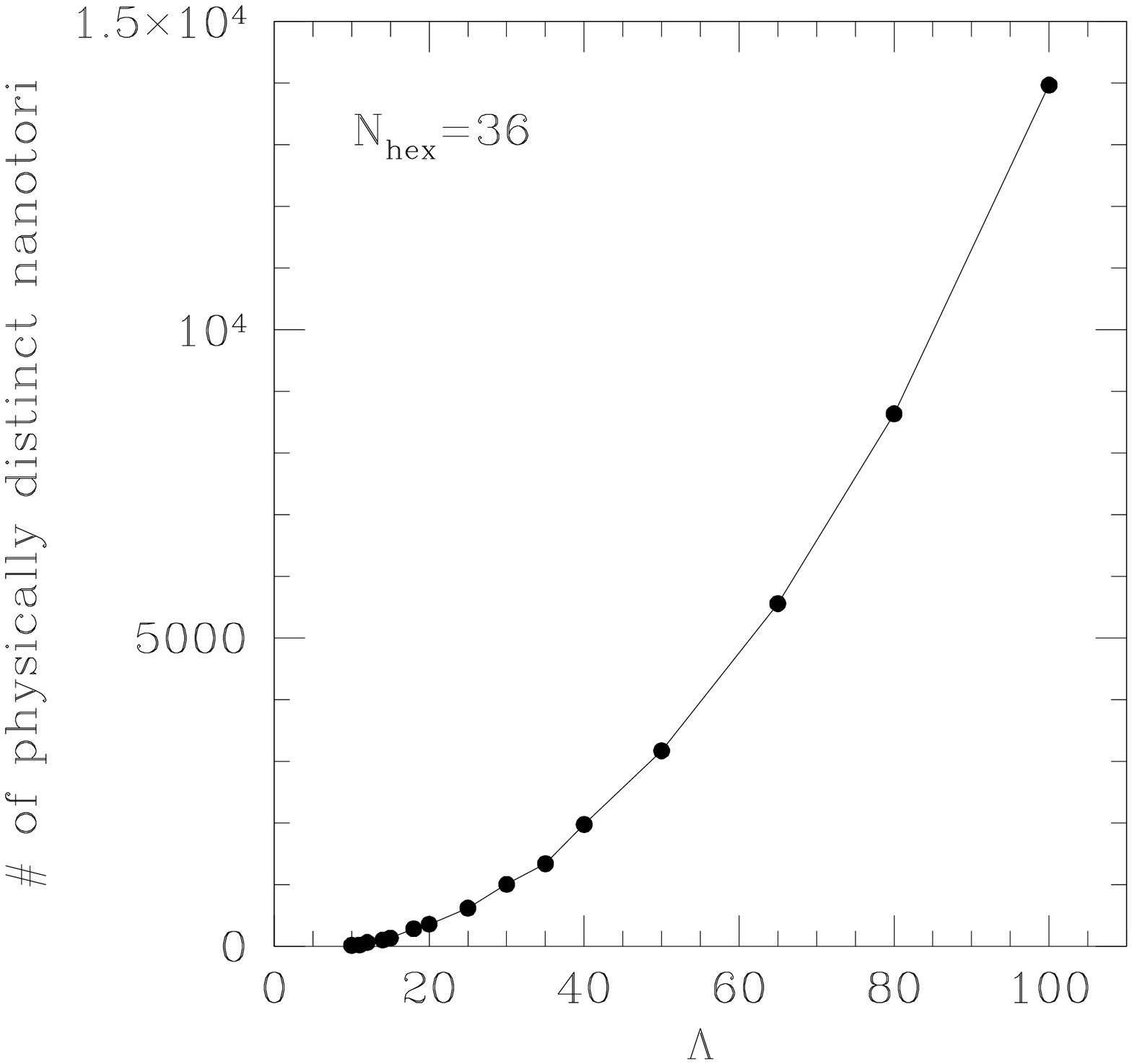} 
   \hskip 0.2truein
   \epsfxsize 3.4 truein \epsfbox{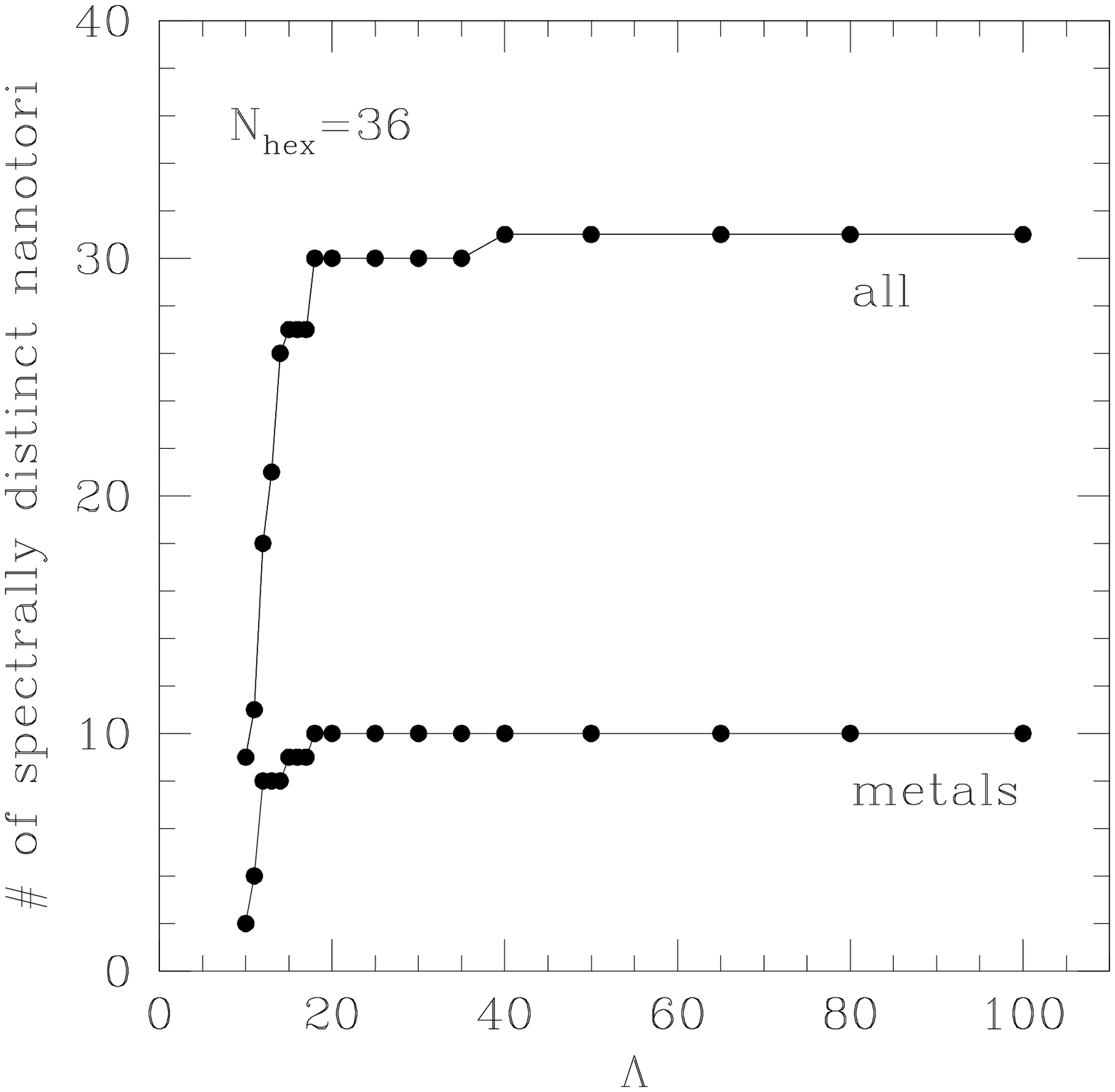} 
 }
\vskip -0.1 truein
\caption{Dramatic reduction in the number of {\it spectrally}\/ distinct carbon nanotori (right panel)
compared with the number of {\it physically}\/ distinct carbon nanotori (left panel).
This illustrates the ubiquity and power of spectral equivalences amongst arbitrary 
sets of allowed nanotori.
Also shown (right panel) is the number of spectrally distinct nanotori which are metallic, indicating 
that metallic properties appear approximately three times more frequently 
amongst {\it spectrally}\/ distinct nanotori than amongst {\it physically}\/ distinct nanotori.}
\label{panels}
\end{figure*}

One might argue that both of these effects --- the huge truncation
in the number of distinct spectral signatures and 
the relative abundance of those which are metallic --- merely reflect the fact
that we restricted our integers $(m,n,p,q)$ to lie within a fixed range bounded by
$\pm \Lambda$, or that we took a relatively small value of $N_{\rm hex}$.
However, it is easy to demonstrate that the second of these effects is relatively 
insensitive to the choice of $\Lambda$, and that the first of these effects only becomes
even more dramatic as $\Lambda$ is increased.
For example, if we restrict our attention to carbon nanotori with 
$N_{\rm hex}=36$, we find that the number of physically distinct
nanotori and the number of spectrally distinct nanotori both rise as a function of $\Lambda$.
However, we see from Fig.~\ref{panels} (left panel) that the number of physically distinct nanotori
with fixed $N_{\rm hex}$ grows as  $\Lambda^2$, as expected, while the number 
of spectrally distinct nanotori quickly hits a plateau (right panel) which remains flat for
an increasingly long interval in $\Lambda$ before a new, hitherto-unrealizable spectral 
signature becomes possible and a new plateau develops.
[This growth in the number of physically distinct nanotori as a function of $\Lambda$ can
easily be deduced from the observation that the number of quadruplets of integers $(m,n,p,q)$
grows as $\Lambda^4$, while the corresponding number of attainable
values of $N_{\rm hex}=np-mp$ grows as $\Lambda^2$.]
As a result, the number of physically distinct nanotori quickly 
outpaces the number of spectrally distinct nanotori.
Moreover, we see from Fig.~\ref{panels} that the number of distinct metallic spectral signatures 
remains roughly one third (and not one ninth) the total number of distinct spectral signatures.

\begin{figure}[h!]
\centerline{
   \epsfxsize 3.25 truein \epsfbox{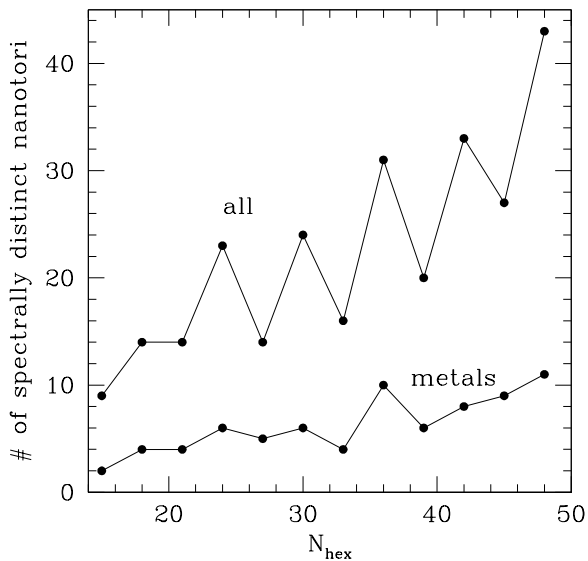} 
 }
\caption{The numbers of spectrally distinct carbon nanotori (upper curve) and metallic spectrally
       distinct carbon nanotori (lower curve),
   plotted as functions of $N_{\rm hex}\in 3\IZ$ for $\Lambda=100$. 
   Both numbers remain
   relatively small.  The ``oscillating'' nature of these
   plots reflects in part the importance of the twist angle $\theta$, since  only non-rectangular 
   tori can exist when $N_{\rm hex}$ is odd.}
\label{Nhexplot}
\end{figure}

We may also consider how
these results vary with the choice of $N_{\rm hex}$.
This is shown in Fig.~\ref{Nhexplot} where, as functions
of $N_{\rm hex}$ and for $\Lambda=100$, we have plotted the number of
spectrally distinct carbon nanotori
as well as the number of those spectrally distinct carbon nanotori
which are metallic.
It is clear from these results that the numbers of spectrally distinct nanotori
remain relatively small,
even though they rise with $N_{\rm hex}$, as expected.

It is also easy to understand the jagged, oscillating nature of the
results in Fig.~\ref{Nhexplot}.
Nanotori must have $N_{\rm hex}\in 3\IZ$ in order to be metallic,
while they must have $N_{\rm hex}\in 2\IZ$ if they are rectangular, with $\theta=\pi/2$. 
Thus, one difference between tori with $N_{\rm hex}\in 6\IZ$ and those with
$N_{\rm hex}\not\in 6\IZ$ is the existence of additional non-rectangular carbon 
nanotori with $\theta\not= \pi/2$.
However, it must also be borne in mind that a random selection of four integers $(m,n,p,q)$
is $5/3$ times more likely to result in an even value for $N_{\rm hex}=np-mq$ than an odd one.
As we see from Fig.~\ref{Nhexplot}, 
the combined effect from these two features is fairly significant.

Finally, another important implication of 
these spectral equivalences between physically distinct carbon nanotori
concerns the traditional rules of thumb which allow us to determine whether 
a given carbon nanotorus is metallic, semiconducting, or insulating.
Throughout the existing literature on this topic, one finds what we shall call the
``rule of three'':  a given $(m,n,p,q)$ carbon nanotorus with $L_2\gg L_1\gg 1$ 
will be metallic if both $m-n$ and $p-q$ are multiples of three,
semiconducting if $m-n$ is multiple of three while $p-q$ is not,
and insulating in all other cases.
It is, of course, easy to verify that this characterization of a metallic
nanotorus is modular invariant.
Specifically, if a given nanotorus has $(m,n,p,q)$ parameters which
are metallic according to the rule of three, then all other nanotori
to which it is spectrally equivalent will also have parameters which
are metallic according to the rule of three.  
In other words, if $m-n$ and $p-q$ are both multiples of three,
modular transformations of these parameters will not disturb this property.

By contrast, the rule-of-three definition of semiconductors 
is {\it not}\/ modular invariant  --- even when we 
preserve the condition that $L_2\gg L_1$.
As a graphic illustration of this point, consider the $(3,0,20,21)$ torus.
Note that indeed $L_2\gg L_1$ for this torus.  According to the standard rule 
of three, such a torus can be identified as a semi-conductor because 
$m-n$ is a multiple of three. 
By contrast, let us now consider the $(23,21,2320,2121)$ nanotorus.
This torus also clearly has $L_2\gg L_1$.
However, because $m-n$ is not a multiple of three, we would
expect this nanotorus to be an insulator.
Indeed, the rule of three tells us to expect this even though our 
first impression might be that the second nanotorus has larger radii
in both directions, and therefore might have energy levels which are
more closely spaced.

However, even though these tori have very different physical parameters,
it turns out that they are related through modular transformations and 
therefore have identical energy spectra.  They therefore also have identical metallicity properties.
This provides a graphic illustration that as a mathematical statement, the 
standard ``rule of three'' fails to characterize the conductivity properties of such
nanotori because it is inconsistent with the modular transformations which 
reflect the additional symmetries of the compactified graphene sheet. 
In other words, the standard definition for a semiconducting carbon nanotorus
fails to be modular invariant, and thus cannot be complete as a description
of the underlying conductivity properties of the nanotorus.
This is yet another consequence of the fact that modular invariance in this
context is an actual physical symmetry relating the spectra of physically
distinct nanostructures.

We close this section with an important comment.
Throughout this section,
our goal has been to illustrate various 
mathematical ramifications of the modular symmetries which govern 
the spectra of different carbon nanotori.
The specific examples we have provided throughout this section 
were therefore chosen for their mathematical simplicity as opposed to their
phenomenological practicality.
For example, we restricted our attention in this section to nanotori
with relatively small values of $N_{\rm hex}$, while
realistic carbon nantoroi can be expected to have 
$N_{\rm hex}\approx {\cal O}(10^2-10^3)$.
Likewise, realistic carbon nanotori will generally be quite long
and thin, with quantities such as $L_2\sin \theta$ exceeding $L_1$
by an order of magnitude or more.
However, all of the conclusions we have drawn in this section
continue to hold even when more realistic tori are considered.
As an example, let us restrict our attention to carbon nanotori 
with $N_{\rm hex}=600$ for which $L_2\sin\theta\geq 10 L_1$.
The latter condition guarantees that our nanotorus remains relatively long and thin
even if there are multiple windings of the hexagonal carbon lattice around
the tube axis of the torus.   
Taking $\Lambda=400$, we find there are $15\,027$ physically distinct nanotori
which satisfy these conditions, but only $52$ of these are spectrally distinct.  Furthermore, of these
$52$ spectral equivalence classes, $14$ correspond to metals.  
We see, then, that even for realistic nanotori,
our modular symmetries continue to lead to large classes of spectrally
equivalent nanotori and a relative overabundance of classes which are metallic.
Indeed, in the limit $\Lambda\to \infty$, it is straightforward to show that the
number of physically distinct carbon nanotori in each spectral 
equivalence class also grows to infinity.

\section{Modular invariance and magnetic fluxes}

With an eye towards potential implications of modular symmetries for the
magnetic phenomena associated with twisted 
carbon nanotori~\cite{MarganskaSzopa,SasakiKawazoe},  
we now consider the introduction of magnetic fluxes.
For full generality, we consider the possibility of {\it two}\/ distinct
fluxes:  one which travels all the way around (and through) the length of the 
nanotube which forms the torus;  and another, namely the usual Aharonov-Bohm flux, 
which pierces the plane of the nanotorus, coming up through the donut hole.    
We shall denote these fluxes $\phi_1$ and $\phi_T$ respectively, as their 
associated vector potentials $\vec{A}_1$ and $\vec{A}_T$ lie parallel to
the vectors $\vec V_1$ and $\vec T$ in Fig.~\ref{fig:TorusParam}, 
respectively.  As is typical for such systems, 
we then find that we can incorporate the effects of these fluxes
by keeping our previous band-structure energy function $E({\vec k})$ in
Eq.~(\ref{eq:BandsTightBinding}) unchanged, 
and simply modifying our constraint equations
for $(k_x,k_y)$ in Eqs.~(\ref{eq:tubek}) and (\ref{eq:torik}) 
so that the integers $\ell_i$ are shifted according to
\beq
            \ell_i ~\to~ \ell_i + {\phi_i\over \phi_0}~
\eeq
where $\phi_0$ is the flux quantum and where 
\beq
           \phi_2 ~\equiv~ \phi_T + \tau_1 \phi_1~.
\eeq
Note that it is the possible existence of 
a non-trivial twist angle $\theta$ which
is responsible for the distinction between $\phi_2$ and $\phi_T$.
We then find that the resulting system continues to exhibit
a spectral equivalence under the modular transformations
in Eqs.~(\ref{eq:STonIndices}) and (\ref{eq:STonVars})
as long as we allow $(\phi_1,\phi_T)$ to remain invariant
under the $T$ transformation and to mix
with each other under the $S$ transformation: 
\beq
         S:~~\pmatrix{\phi_1\cr \phi_T\cr}
  \to \pmatrix{\phi'_1\cr \phi'_T\cr} \equiv
       \pmatrix{   -\tau_1 & -1 \cr
                    \tau_2^2/ |\tau|^2 & -\tau_1/|\tau|^2 \cr}
         \pmatrix{\phi_1\cr \phi_T\cr}~.
\eeq
Thus, a carbon nanotorus parametrized by $(m,n,p,q,\phi_1,\phi_T)$
will have the same spectral properties as one parametrized by
$(m',n',p',q',\phi'_1,\phi'_T)$, where these two sets of parameters
are related through the modular transformations discussed above.


\section{Discussion}

In this paper, we have highlighted and investigated the implications
of a geometric symmetry --- modular invariance --- which emerges 
upon the compactification of
a graphene sheet to form a carbon nanotorus.
Although not traditionally considered in the carbon-nanotorus 
literature, modular invariance plays a critical
role in describing the spectral properties of these nanotori
and leads to spectral equivalences between physically distinct
nanotori.  As we have shown,
this has profound implications for the 
classification of carbon nanotori, indicating 
that large numbers of seemingly unrelated nanotori are in 
fact completely identical in terms of their spectral properties.
Along the way, we also showed that  
the traditional ``rule of three'' classification rubric
is incomplete, as it is based on quantities which
are do not respect these spectral equivalences.
We also found that the fraction of spectrally distinct carbon 
nanotori which are metals is approximately three times greater 
than would naively be
expected on the basis of standard results in the literature.
Finally, we also showed that these spectral equivalences can easily be extended
to cases in which non-trivial magnetic fluxes are present.

The existence of these spectral symmetries also provides a deeper
theoretical underpinning to certain results which already exist
in the literature.  For example, 
it is well known that
many carbon nanotori exhibit persistent currents
in the presence of a non-zero magnetic flux $\phi$.
As functions of $\phi/\phi_0$, these currents typically follow
complicated ``sawtooth'' patterns which have a natural periodicity
under shifts $\phi\to\phi+\phi_0$.
It has also separately been observed (see, \eg, Ref.~\cite{SasakiKawazoe})
that any such sawtooth pattern is preserved but shifted horizontally
upon the introduction of a nanotorus twist in which 
$\vec V_2\to \vec V'_2 \equiv \vec V_2+ f \vec V_1$,
where $f$ is chosen such that $\vec V'_2$ is also a lattice vector.
Indeed, when $f\in \IZ$, the magnitude of this horizontal
shift exactly matches the periodicity of the sawtooth 
pattern and the net result is unchanged.

Remarkably, this coincidence is now easy to understand from the point
of view of modular transformations:
when $f\in \IZ$, the mapping from $\vec V_2\to \vec V'_2$
is nothing but the $T$ modular transformation, and as we have shown,
modular transformations
preserve the electrical properties of the torus, including
its persistent currents. 
We thus see that the periodicity of the sawtooth pattern
for persistent currents under shifts $\phi\to\phi+\phi_0$ --- a
periodicity which can be understood on elementary grounds having nothing
to do with modular transformations --- can now also be interpreted as a spectral
equivalence under the $T$ modular transformation.
Moreover, we now see that this is merely the tip of the
iceberg, and that these sorts of spectral equivalences actually have
a richer structure and context that 
not only corresponds to the entire modular group
but also transcends the specific example of persistent currents.

As we have discussed above,
the traditional ``rule of three'' is not formulated in a
modular-invariant way and is therefore incomplete
as a description of the electronic properties of carbon nanotori.
However, the full implications of these modular symmetries are 
significantly broader than just the rule of three:
no theoretical result concerning the electronic properties of
carbon nanotori can be correct unless it respects these modular symmetries.
In other words, no calculation of any electronic property of a given carbon nanotorus
in terms of its fundamental defining parameters $(m,n,p,q)$ can be
correct unless it yields a result which is invariant under the modular
transformations in Eq.~(\ref{eq:STonIndices}).
In this respect, modular invariance functions for carbon
nanotori in much the same way as gauge invariance functions for
electromagnetic systems:  no theoretical result can be correct unless it
can be phrased in terms of quantities which are invariant under the
symmetry in question.  
Of course, at a mathematical level, both modular
invariance and gauge invariance rest on relatively simple algebraic identities.
However, they both provide powerful organizing principles, and 
have significant physical manifestations and
implications for the symmetry structures of the systems in which
they appear.

Needless to say, several additional comments are in order.  First,
it should be noted that 
in this paper we have focused on what might called the ``ideal''
nanotorus.  In particular, we have not accounted for the fact that 
the actual physical construction of such a torus in three-dimensional space
requires that we introduce both an intrinsic and extrinsic curvature onto 
our otherwise flat graphene sheet.
In this sense, the construction of a carbon nanotorus is different
from that of a carbon nanotube (in which only the extrinsic curvature is non-vanishing).
The introduction of intrinsic curvature requires that we subject our underlying
graphene sheet to considerable strain, deforming not only the positions of carbon atoms
but also their relative spacings.  These effects have been addressed 
by a number of authors, using a variety of different 
techniques~\cite{CurvatureDistortion}.

That said, these spectral equivalences should continue to hold,
even in the presence of such deformations.
There are several reasons for this.
First, in the limit $L_1,L_2\gg R_{cc}$,
all effects due to these deformations will be suppressed.
However, this is precisely the limit in which nanotori can be constructed
from purely hexagonal graphene sheets 
without the introduction of curvature-inducing pentagonal or heptagonal carbon rings.
Second, it can be shown that even when such strain is present, maximum toroidal stability
occurs when this strain is
uniformly distributed along the nanotorus~\cite{strain}, and this is precisely
the situation in which the techniques of Ref.~\cite{KaneMele} can be used in order to
mathematically rewrite the effects of such strain as arising due to a fictitious magnetic
flux.  As we have seen, the spectral equivalences we have found 
continue to exist even when such fluxes are introduced.
But most importantly, while {\it any}\/ deformations of the underlying graphene sheet
can be expected to have effects
on the corresponding band-structure energy function $E(\vec k)$ shown 
in Fig.~\ref{fig:EnergyContours}, such deformations will not
disturb the symmetries inherent in the constraint equations for $\vec k=(k_x,k_y)$ 
derived in Eqs.~(\ref{eq:tubek})
and (\ref{eq:tubek2}).  Indeed, these equations reflect nothing more than 
the effects of toroidal compactification, and their structure leads directly
to the modular symmetries inherent in Eqs.~(\ref{eq:torik}) and (\ref{eq:torik2}).
Thus, since these spectral equivalences ultimately stem from these symmetry properties, 
we are assured that any two tori related by modular transformations will sample the same
set of wavevectors $\vec k$.  Such tori will therefore continue to be spectrally identical
regardless of the function $E(\vec k)$, provided the deformations to $E(\vec k)$ for the
two tori are themselves identical.  Moreover, as we have argued above, 
even in cases where these deformations
are not identical, they can at most differ by terms which are suppressed
by factors of the presumably large nanotori radii.  
Such differences can therefore be safely neglected.

A similar conclusion also holds for thermal effects.
It might seem, at first glance, that thermal effects could also destroy the
spectral equivalences, since they too can have a dramatic effect on the band structure
of the underlying graphene sheet~\cite{ThermalDisorder}.  
However, as noted above, these spectral equivalences 
are a consequence of the geometric symmetries 
that arise upon {\it compactifying}\/  this sheet;  they are
largely independent of the symmetries of the sheet itself.   
Indeed, these modular symmetries exist for {\it any}\/
choices of identification vectors $\vec V_1$ and $\vec V_2$, even if
those vectors are altered by other effects. 
Therefore, as long as the temperature is sufficiently low that the electron coherence length 
exceeds both the inner and outer toroidal circumferences,
the spectral equivalences we have been discussing should remain intact up to terms
suppressed by the large nanotori radii.

These comments notwithstanding, it still remains true that not all such
nanotori are equally likely to appear in nature.  
In particular, those nanotori whose constructions implicitly involve large numbers of twists are
likely to be rather difficult to construct or stablize, as the carbon atoms
that constitute the underlying graphene sheet are likely to experience significant strain,
leading to major deformations of the underlying carbon bond lengths and angles away from their
ideal values. 
This is especially
relevant, given that our statistical discussions in Sect.~IV implicitly assume 
that each of the relevant $(m,n,p,q)$ nanotori can appear with equal probability.

At a mathematical level, it is difficult to draw a firm boundary between those 
carbon nanotori which are realizable in nature as {\it bona-fide}\/ molecules and those which are not.
However, in this paper we have restricted ourselves
to statistical examinations of only those nanotori which already obey 
certain critical constraints.  For example, we have limited ourselves to nanotori in 
which $L_1$ and $L_2$ are both significantly bigger than the carbon-carbon bond length $R_{cc}$.
Thus each twist experienced for the whole nanotorus 
has only a minor effect on the bond lengths and angles corresponding
to each individual carbon atom.
In addition, we have further limited ourselves to nanotori which also satisfy $L_2\gg L_1\sin \theta$.
This ensures that each such nanotorus can be realized in three-dimensional space,
without the unphysical self-overlapping that would arise if this condition
were not met.
Together, these conditions help to ensure that the strains induced by these modular transformations
are not too severe.

    
Needless to say, there are many implications of these spectral
equivalences which we have not yet explored.
It would be interesting, for example, to consider the implications
of these symmetries for the existence or absence of
persistent currents as well as the existence or absence of colossal magnetic moments.
This work is currently in progress.
It would also be interesting to understand these modular transformations 
in terms of fictitious fluxes, using analogues of the techniques presented
in Ref.~\cite{KaneMele}. 

In closing, we would like to make two final remarks, one of
primarily mathematical interest and one of more practical applicability.

First, as we have seen, modular invariance is the symmetry which underlies
most of the results we have presented in this paper.  At a mathematical level,
modular invariance is normally just a relabelling
symmetry in the sense that
two sets of torus parameters related by a modular transformation 
normally correspond to the same physical torus.
This is certainly the case in string theory, and 
in most other situations in theoretical high-energy physics
in which modular transformations have played
a significant role (see, \eg, Ref.~\cite{shape}). 

However, the case of carbon nanotori
is quite different.  Here, 
modular transformations relate parameters corresponding to 
carbon nanotori
which are physically distinct.  As we have discussed, this is because  
we are not merely rolling up an unmarked sheet of paper when
we subject it to two non-parallel identifications;  we are rolling
up a graphene sheet which already has a hexagonal carbon lattice imprinted on it.
Viewed from this perspective, 
it is therefore somewhat remarkable that modular 
transformations continue to play a role, indicating
when two distinct nanotori will have the same spectral properties.
Indeed, in the case of carbon nanotori, 
we see that modular invariance is thus promoted
from a mere relabeling symmetry to something far deeper:  
 {\it modular invariance becomes an outright physical symmetry 
between physically distinct entities}.
We know of no other physical situation in which modular invariance plays such a role. 

Second, it is also exciting at a practical level that
physically distinct carbon nanotori
can have identical energy spectra and electrical properties.
Since these nanotori are physically distinct, some are likely to be far more 
complicated to construct in the laboratory than others.
Nevertheless, these spectral equivalences suggest that it may not be necessary 
to fabricate a very complex nanotorus (or generate sizable magnetic fluxes)
in order to obtain a desired spectral property;
there are likely to be far simpler nanotori and/or fluxes which can perform the same function. 
This could have significant implications for the production and use of such 
nano-materials.


\smallskip
This work was supported in part
by the Department of Energy under Grant~DE-FG02-04ER-41298.
We are happy to thank P.~McEuen, D.~Ralph, C.~Stafford, and H.~Tye for discussions,
and we wish to acknowledge the hospitality of the Kavli Institute for Theoretical
Physics (KITP) in Santa Barbara, California, where this paper was completed.
The opinions and conclusions expressed here are those of the authors,
and do not represent either the Department of Energy or the National Science Foundation.


\medskip


\end{document}